\documentclass[twocolumn]{aastex62}
\usepackage{natbib}
\usepackage{amsmath}
\submitjournal{ApJ}

\shorttitle{SkyMapper meets \textit{Gaia} DR2 to explore the Milky Way halo}
\shortauthors{C. L. Sahlholdt et al.}

\begin{document}

\title{Characteristics of the two sequences seen in the high-velocity Hertzsprung--Russell diagram in \textit{Gaia} DR2\footnote{Based on data from the SkyMapper Southern Sky Survey}}

\correspondingauthor{Christian L. Sahlholdt}
\email{christian.sahlholdt@astro.lu.se}

\author[0000-0002-0918-346X]{Christian L. Sahlholdt}
\affil{Lund Observatory, Department of Astronomy and Theoretical Physics, Box 43,
      SE-221 00 Lund, Sweden}

\author[0000-0003-2688-7511]{Luca Casagrande}
\affil{Research School of Astronomy \& Astrophysics, Mount Stromlo Observatory,
      The Australian National University, ACT 2611, Australia}

\author[0000-0002-7539-1638]{Sofia Feltzing}
\affil{Lund Observatory, Department of Astronomy and Theoretical Physics, Box 43,
      SE-221 00 Lund, Sweden}



\begin{abstract}
In this study we use a sample of about 9 million SkyMapper stars with metallicities to investigate the properties of the two stellar populations seen in the high-velocity ($V_{\rm T} > 200$\,km\,s$^{-1}$) {\it Gaia} DR2 Hertzsprung--Russell diagram.
Based on 10,000 red giant branch (RGB) stars (out of 75,000 with high velocity), we find that the two sequences have different metallicity distribution functions; one peaks at $-1.4$~dex (blue sequence) and the other at $-0.7$~dex (red sequence).
Isochrones with ages in the range $11$--$13.5$~Gyr, and metallicities chosen to match the observations for each sequence, fit the turnoffs and broad RGBs well, indicating that the two populations formed at comparable times within the uncertainties.
We find that the mean tangential velocity of disk stars increases steadily with decreasing metallicity, and that the red sequence is made up of the high-velocity stars at the lowest metallicities of the thick-disk population.
Using relative number densities, we further find that the red-sequence stars are more centrally concentrated in the Galaxy, and we estimate the radial scale length of this population to be on the order of 2--3~kpc.
The blue-sequence stars, on the other hand, follow a nearly flat radial density profile.
These findings tighten the link between the red-sequence stars and the chemically defined thick disk.
\end{abstract}



\section{Introduction} \label{sect:intro}

The second data release by \textit{Gaia} (DR2) showed that stars with tangential velocities, $V_{\rm T}$, greater than $200$~km\,s$^{-1}$ fall along two distinct, well-defined sequences in the Hertzsprung--Russell (H-R) diagram \citep[][their figure~21]{2018A&A...616A..10G}.
The two sequences are offset by about 0.1~mag in color and are commonly referred to as the blue and red sequences.
\citet{2018A&A...616A..10G} suggested that the two sequences could tentatively be associated with the two groups of halo stars with distinct elemental abundance trends seen in \citet{2010A&A...511L..10N}.
Additionally, a population of stars on radial and slightly retrograde orbits identified in kinematical studies of halo stars even before \textit{Gaia} DR2 \citep{2018MNRAS.478..611B} has been found to fall along the blue sequence \citep{2018ApJ...860L..11K}.
Using elemental abundances for stars in this population \citep[from APOGEE;][]{2018ApJS..235...42A}, \citet{2018Natur.563...85H} argue that it is made up of debris from a galaxy accreted onto the Milky Way.
\citet{2018ApJ...863..113H} share this view and showed that APOGEE stars falling on the red sequence have elemental abundances consistent with the chemically defined thick disk.
They further suggested that this high-velocity thick-disk component could have been heated by the merger that brought in the stars on the blue sequence \citep[see also][]{2018arXiv181208232D}.

\citet{2019arXiv190102900G} modeled the full {\it Gaia} H-R~diagrams for halo and thick-disk stars, with halo stars defined as those with $V_{\rm T} > 200$~km\,s$^{-1}$ and thick-disk stars defined as having $V_{\rm T} < 200$~km\,s$^{-1}$ and a distance from the Galactic plane of more than 1.1~kpc.
They find that they are able to faithfully reproduce the H-R~diagram for the halo with two stellar populations, each with the same range of ages but different metallicity distributions.
However, they interpret the red sequence as being the in-situ halo since it is older than what they define as the thick disk.

Thus, the overall picture that has emerged is one in which the local Galactic halo (defined as stars with $V_{\rm T} > 200$~km\,s$^{-1}$) consists of two coeval populations with different metallicity distributions.
The more metal-poor population makes up the blue sequence and is thought to consist of debris from a single significant accretion event.
The more metal-rich population that makes up the red sequence may be the in-situ halo population or the high-velocity tail of the thick disk heated by mergers.
However, we note that although \textit{Gaia} gives good distances, magnitudes, and colors, the inference about the metallicity distributions in the studies cited above is based on high-quality, but in numbers really rather limited, data sets from, e.g. \citet{2010A&A...511L..10N} and APOGEE DR14 \citep{2018ApJS..235...42A}.

Recently, \citet{2019MNRAS.482.2770C} have determined spectroscopically calibrated metallicities for on the order of 10 million stars observed in the SkyMapper Southern Sky Survey and included in {\it Gaia} DR2.
With these data we have metallicity estimates for a largely unbiased sample of about 75,000 high-velocity stars with high-quality astrometry.
In this paper we use this new extensive data set to provide additional perspectives on the two  high-velocity populations as well as their link to the halo and disk populations.
Following our analysis, we discuss the results in light of the picture discussed above.

\section{Data} \label{sect:data}
 
\subsection{Stellar parameters}\label{sect:params}

For our investigation we take metallicities ([Fe/H]) from \citet{2019MNRAS.482.2770C} who used SkyMapper photometry to derive metallicities as well as temperatures for $\sim$10 million stars cross-matched with \textit{Gaia} DR2 \citep{2018A&A...616A...1G}. 

We are confident that these metallicities are good across the red giant branch (RGB); however, at the turnoff our investigations show that the SkyMapper calibration has some limitations, and metallicities below about $-1.5$\,dex cannot be derived with confidence. 
Our analysis show that it is likely the small range of stellar parameters in the GALAH calibrating data set that has led to the poor quality of metallicities for turnoff stars below  $-1.5$\,dex.
With GALAH DR3 on the horizon the SkyMapper team is set to update their calibration; however, we do not need to wait for this to undertake the current investigation, as by looking at the \textit{Gaia} colors and magnitudes and using the knowledge about metallicities from the RGB as a prior to our investigation, we can circumnavigate this particular problem.
 
The SkyMapper magnitudes have been corrected for reddening using rescaled $E(B-V)$ from the \citet{1998ApJ...500..525S} map as described by \citet{2019MNRAS.482.2770C}.
We use these same $E(B-V)$ estimates, combined with the mean extinction coefficients given by \citet[][Table 2]{2018MNRAS.479L.102C}, to correct the {\it Gaia} magnitudes for reddening.
We follow \citet{2018A&A...616A..10G} to calculate tangential velocities for the stars based on \textit{Gaia} DR2 parallaxes and proper motions and will refer to those with $V_{\rm T} > 200$ km\,s$^{-1}$ as high-velocity stars \citep[compare][Figure 21]{2018A&A...616A..10G}.
When calculating absolute magnitudes, we apply a zero-point correction of 0.03~mas to the parallaxes \citep{2018A&A...616A...2L}.
We do not apply this offset before calculating $V_{\rm T}$ to stay consistent with \citet{2018A&A...616A...1G}, but we find that this choice does not affect our conclusions.

\begin{figure}
\centering
\includegraphics[width=\columnwidth]{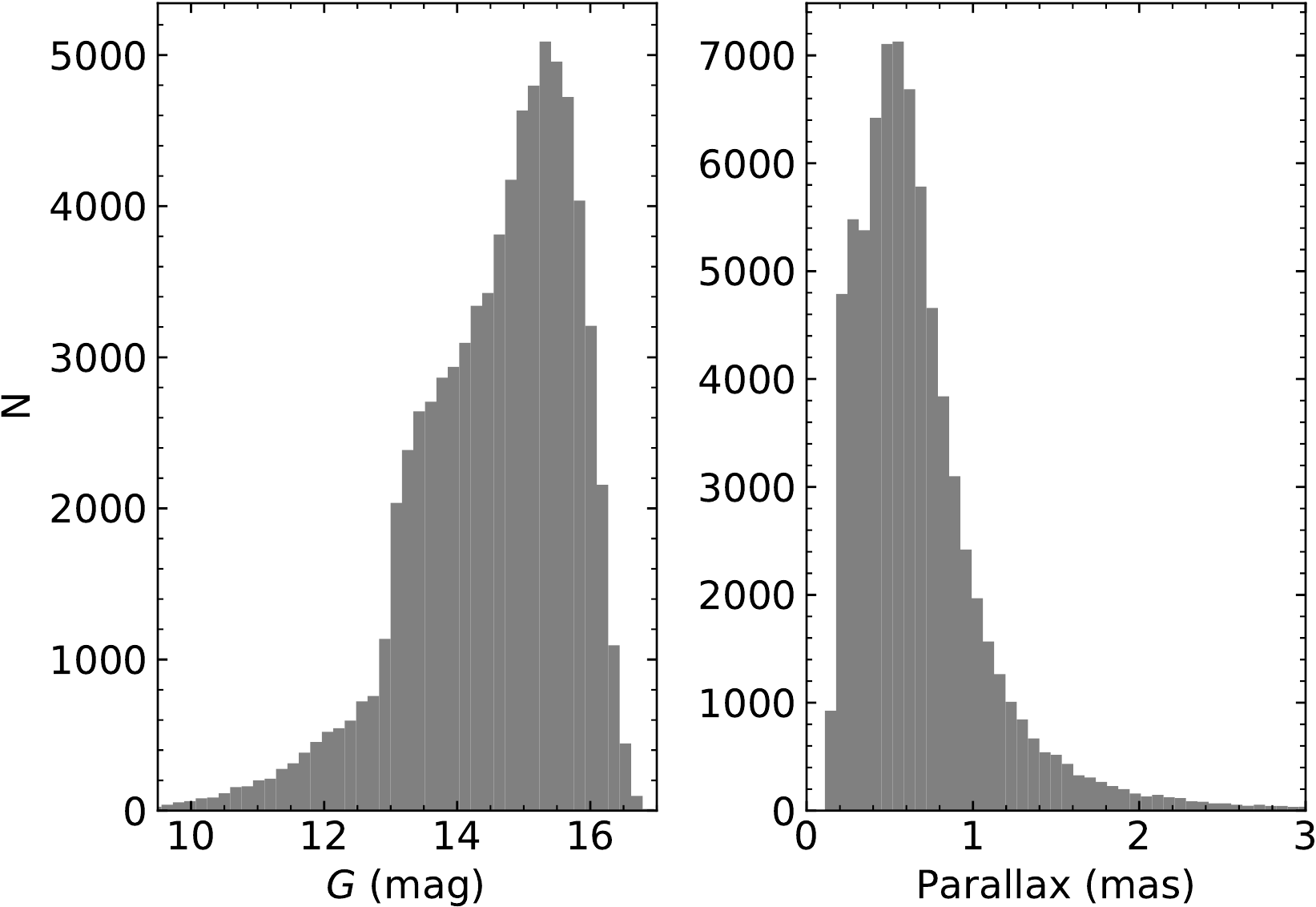}
\caption{Distributions of apparent $G$ magnitude and parallax for the high-velocity ($V_{\rm T} > 200$\,km\,s$^{-1}$) SkyMapper sample.}
\label{fig:mag_plx_hist}
\end{figure}

\subsection{Quality cuts}
We trim the data set by applying the constraints from \citet[][Equations~(1) and (3)]{2018A&A...616A..17A} that are designed to filter out spurious astrometric solutions in the {\it Gaia} data.
The constraint from their Equation~(2) has not been applied since that is based on photometric quality (avoiding problems in crowded fields), and our stars have already been culled to not have a source closer than 15$''$ from SkyMapper photometry \citep{2019MNRAS.482.2770C}.
Additionally, we constrain the relative uncertainty on the parallax to be no higher than 10\%.
These cuts leave a total of about $5.3$ million stars of which 75,000 have $V_{\rm T} > 200$\,km\,s$^{-1}$.
Fig.~\ref{fig:mag_plx_hist} shows the distributions of apparent magnitude and parallax for the high-velocity stars.
Our main focus here is on this high-velocity subsample.

\section{Analysis}

\subsection{Metallicities of the stellar sequences}

\begin{figure}
\centering
\includegraphics[width=\columnwidth]{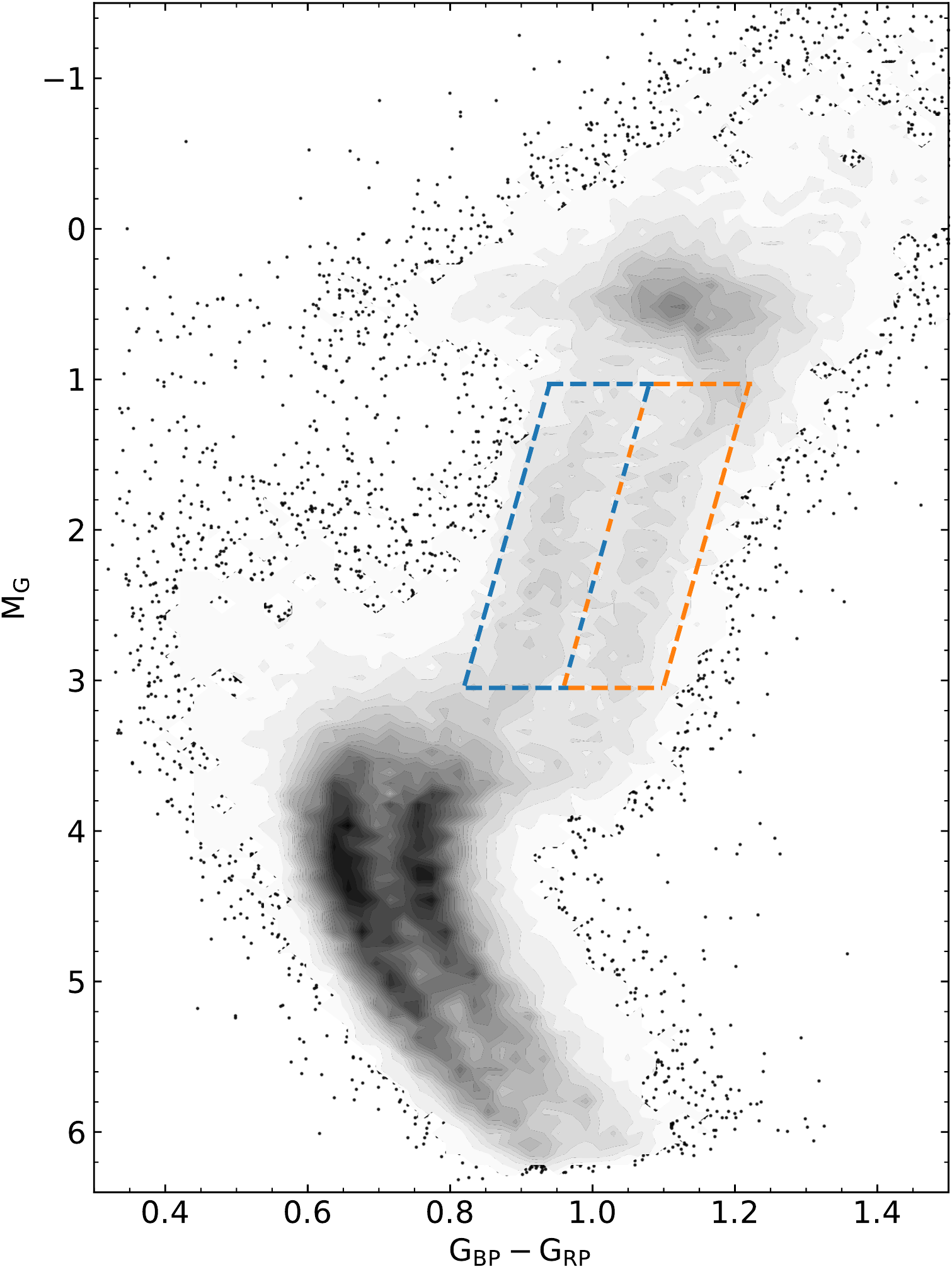}
\caption{H-R diagram of the SkyMapper high-velocity stars.
The data are plotted as density contours except in low-density regions.
Two sequences are visible, most clearly at the turnoff and the RGB.
The blue and orange boxes indicate our selection of blue- and red-sequence stars on the RGB.}
\label{fig:hrd1}
\end{figure}

\begin{figure}
\centering
\includegraphics[width=\columnwidth]{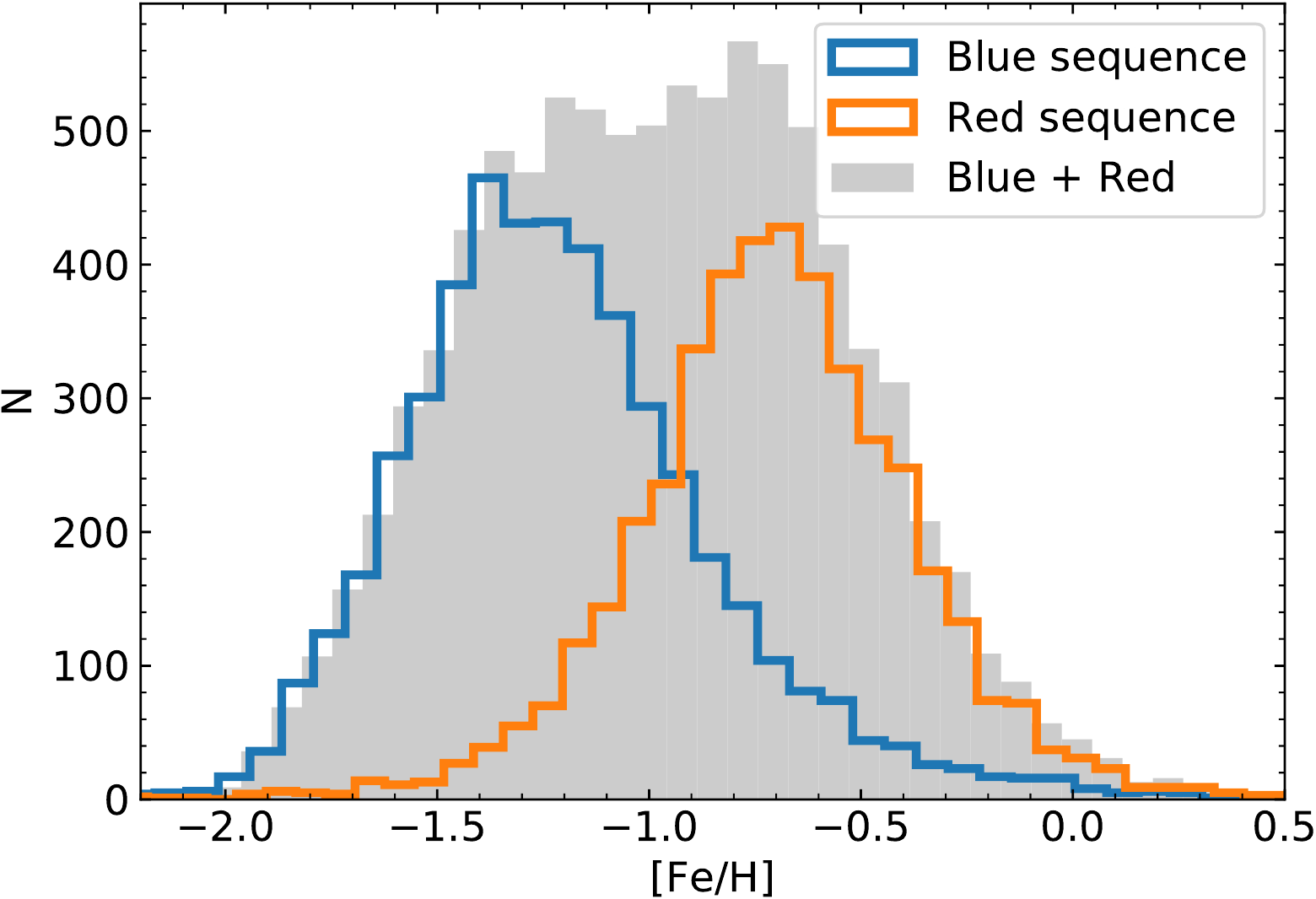}
\caption{Metallicity histograms for the SkyMapper stars falling in the parallelograms drawn on the RGB in Fig.~\ref{fig:hrd1}.
The blue and orange histograms are for stars in the left- and right-hand parallelograms, respectively, and the gray histogram is the combination of the two.}
\label{fig:met_hist}
\end{figure}

Our H-R diagram for the SkyMapper high-velocity stars is shown in Fig.~\ref{fig:hrd1}.
The stars are clearly separated into blue and red sequences both at the turnoff and along the RGB \citep[compare][their figure~21]{2018A&A...616A..10G}.

If we analyze the stars on the RGB, we find that the metallicity distribution function (MDF) shows a plateau for metallicities between $-1.5$ and $-0.5$~dex (Fig.~\ref{fig:met_hist}, gray histogram).
When splitting the MDF into two, by picking out stars along the two sequences in the H-R diagram (colored boxes in Fig.~\ref{fig:hrd1}), the plateau is found to be the combination of two broad MDFs peaking at $-1.4$ (blue sequence) and $-0.7$~dex (red sequence), respectively.
There is roughly an equal number of stars in the two sequences.
The two MDFs are approximately Gaussian with standard deviations of 0.36 (blue) and 0.33~dex (red).
For comparison, the mean uncertainty on the metallicities is 0.18~dex, so the spread of the distributions is not purely driven by the uncertainties.
Our simple split of the sequences likely leads to misclassification of some of the stars which contributes to the tails of the MDFs (more high metallicities in the blue sequence and low metallicities in the red sequence).
Still, we can confidently say that the two sequences are to first order driven by a difference in metallicity.

\subsection{Turnoff ages of the sequences}

\begin{figure}
\centering
\includegraphics[width=\columnwidth]{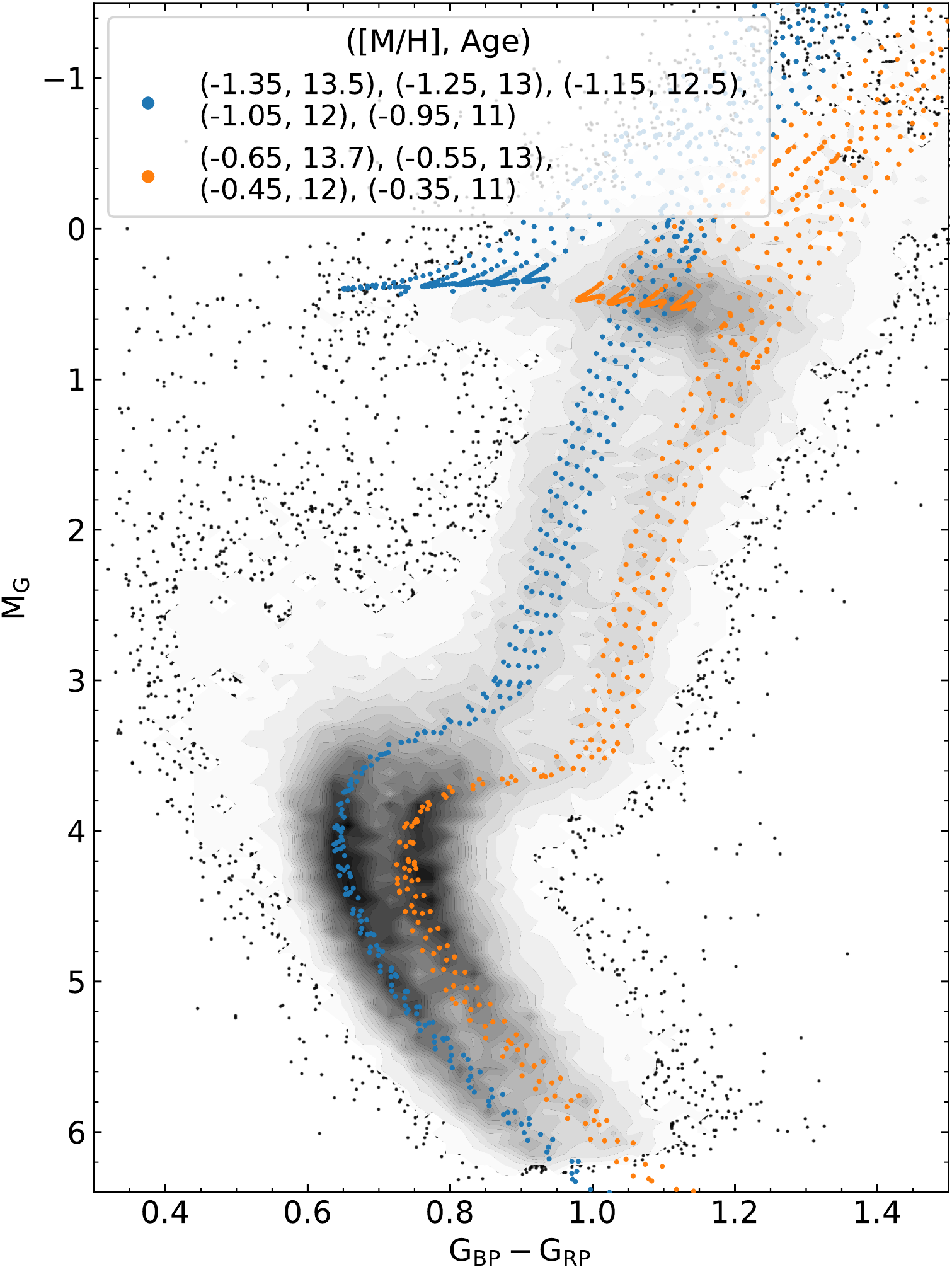}
\caption{H-R diagram of the SkyMapper high-velocity stars like in Fig~\ref{fig:hrd1}.
Isochrones from the PARSEC database with a range of metallicities and ages, with metallicities chosen to match the observations on the giant branch, are plotted on top of the data (see the legend).}
\label{fig:hrd2}
\end{figure}

In Fig.~\ref{fig:hrd2} we further explore the nature of the sequences in the H-R diagram.
The turnoffs of the two sequences are very sharply defined, and now that we know the metallicities of the stars in the two sequences, we can utilize these sharp turnoffs to constrain their ages by using isochrones with the relevant metallicities.
We use PARSEC isochrones \citep{2012MNRAS.427..127B} with {\it Gaia} DR2 passbands from \citet{2018A&A...619A.180M}.
Since we do not know the $\alpha$-enhancement of the stars, we have used isochrones with solar-scaled abundances and the \citet{1993ApJ...414..580S} prescription for scaling the metallicity with $\alpha$-enhancement assuming [$\alpha$/Fe]\,$=0.2$ which is appropriate if the two sequences are made up of disk and accreted halo stars.
This means that if, for example, the data indicate $[\mathrm{Fe}/\mathrm{H}]=-1.5$, we use their formula to get $[\mathrm{M}/\mathrm{H}]\approx -1.35$ and pick an isochrone with this value for [Fe/H].
For each metallicity, we adjust the age of the isochrone (by eye) such that it follows the turnoff.
We repeat this exercise for each of the sequences with metallicities chosen around the maxima of the MDFs in Fig.~\ref{fig:met_hist}.
In the end, the blue isochrones in Fig.~\ref{fig:hrd2} have ages from $11$ to $13.5$~Gyr and the red ones from $11$ to $13.7$~Gyr.

This simple exercise indicates that the same range of ages for the two sequences can explain the data, and that they both formed early in the history of the Milky Way in environments of different metallicities.
However, our age estimates are somewhat uncertain given that they are based on fits by eye, and an increase of the $\alpha$-enhancement to [$\alpha$/Fe]\,$=0.4$ would lower the age estimates by about 1 Gyr.
Thus, the star formation histories of the two sequences may not overlap exactly as implied by our estimates, but given the spread of about 3~Gyr in the ages that fit each sequence, it is likely that at least some of their stars were formed simultaneously.
It is also interesting to note that we match the sharp turnoffs by picking older isochrones at lower metallicities because the increase in the turnoff temperature at lower metallicities is compensated by a decrease in the temperature due to the higher age.
On the RGB, however, the age does not affect the temperature significantly, and isochrones of different metallicities describe the spread observed in each sequence along the RGB.

\subsection{Tangential velocity distributions of the sequences}

\begin{figure}
\centering
\includegraphics[width=\columnwidth]{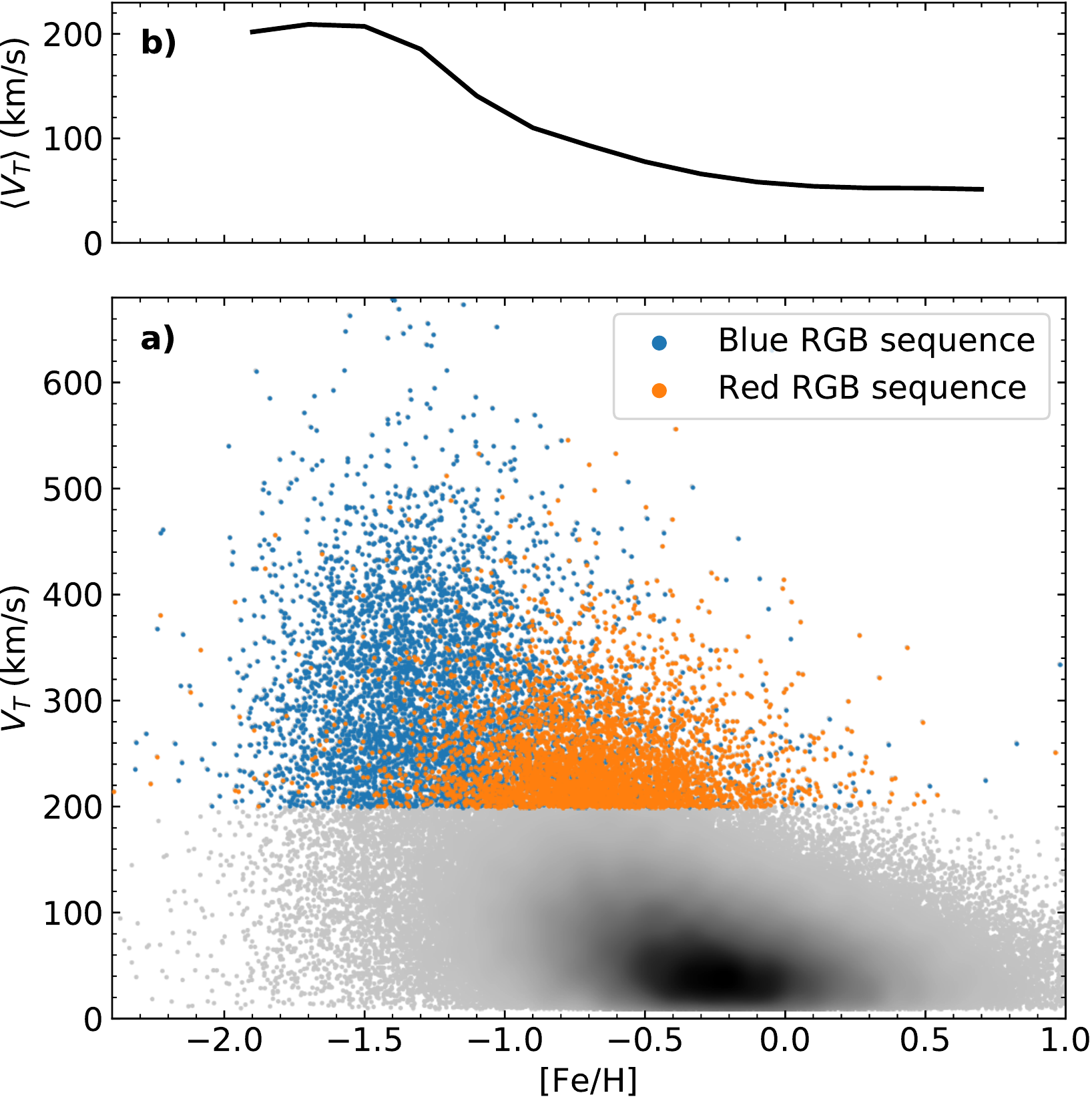}
\caption{(a) $V_{\rm T}$ vs. [Fe/H] for stars on the red giant branch selected as shown in Fig.~\ref{fig:hrd1}.
(b) Mean $V_{\rm T}$ as a function of metallicity.}
\label{fig:vtan_feh}
\end{figure}

Next, we explore the distribution of tangential velocities for the two sequences in the H-R diagram and find that we confirm the tentative conclusion by \citet{2018A&A...616A..10G}, namely, that the blue and more metal-poor sequence spreads to higher $V_{\rm T}$ than the red sequence (Fig.~\ref{fig:vtan_feh}a).
There is a very clear split between the two sequences in the distributions of $V_{\rm T}$; on the red sequence, very few stars reach $V_{\rm T} > 300$ km\,s$^{-1}$, which is not uncommon on the blue sequence.
So almost all of the stars with the highest $V_{\rm T}$ are metal-poor, typically with [Fe/H]~$< -1.0$~dex.

The high-velocity stars of the red sequence seem to be just the tip of the population of disk stars lying mainly at lower velocities.
This population of stars shows a steady increase in the maximum value of $V_{\rm T}$ as the metallicity decreases.
In contrast, the blue-sequence stars have a very similar distribution in $V_{\rm T}$ at all metallicities where it is present.
This is reflected in the flattening of the mean $V_{\rm T}$ at low metallicities (Fig.~\ref{fig:vtan_feh}b) where the blue sequence dominates.

\subsection{The red sequence and the thick disk}
\label{sect:thick_disk}

\begin{figure*}
\centering
\includegraphics[width=\textwidth]{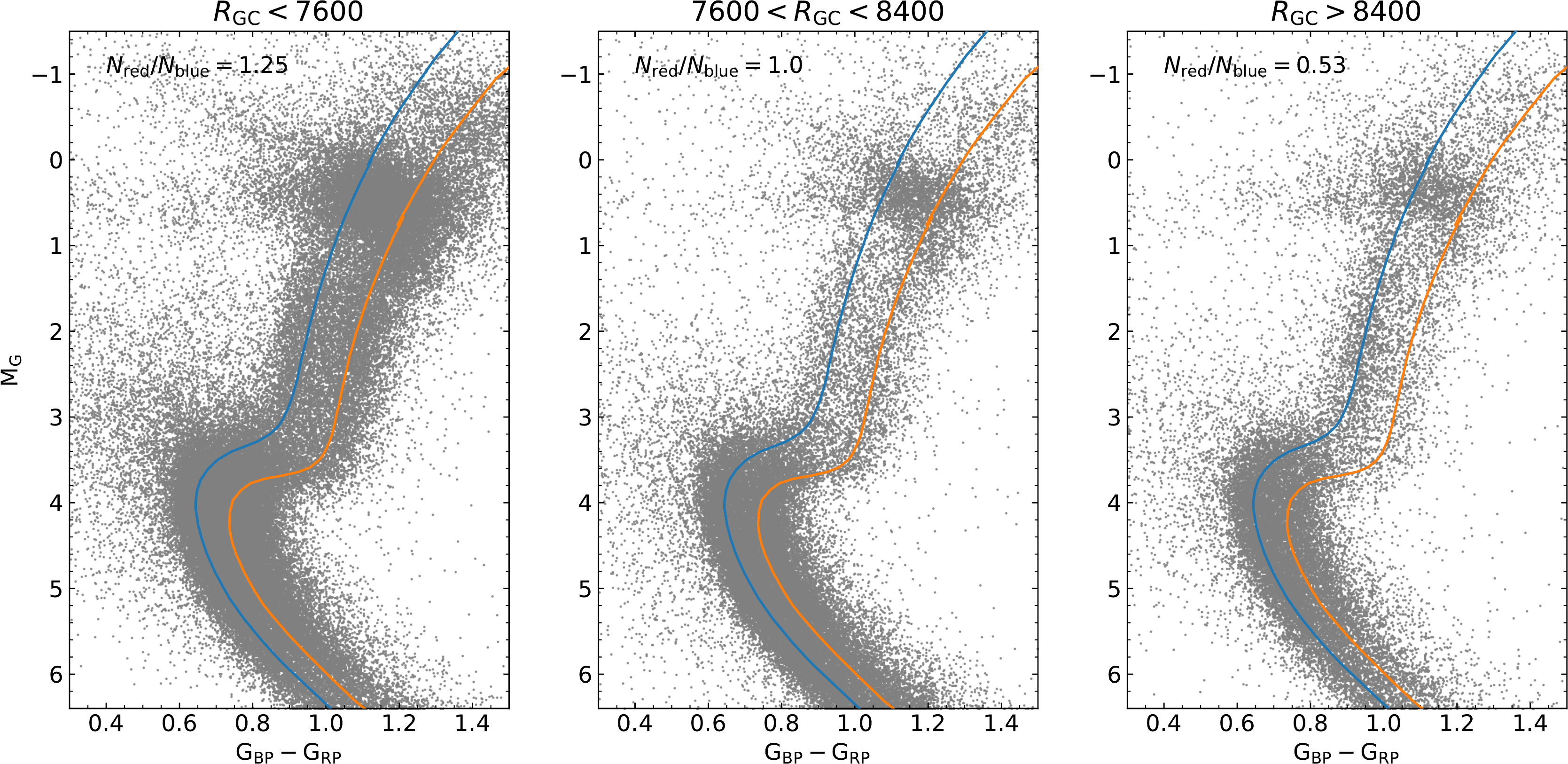}
\caption{H-R diagrams of  {\it Gaia} high-velocity stars in different intervals of galactocentric radius. The two isochrones are shown as guides tracing out the two sequences.
$N_{\mathrm{red}}/N_{\mathrm{blue}}$ is the fraction of red-to-blue sequence stars on the RGB counted within 0.05~mag in color around the isochrones.}
\label{fig:hrd3}
\end{figure*}

\begin{figure}
\centering
\includegraphics[width=\columnwidth]{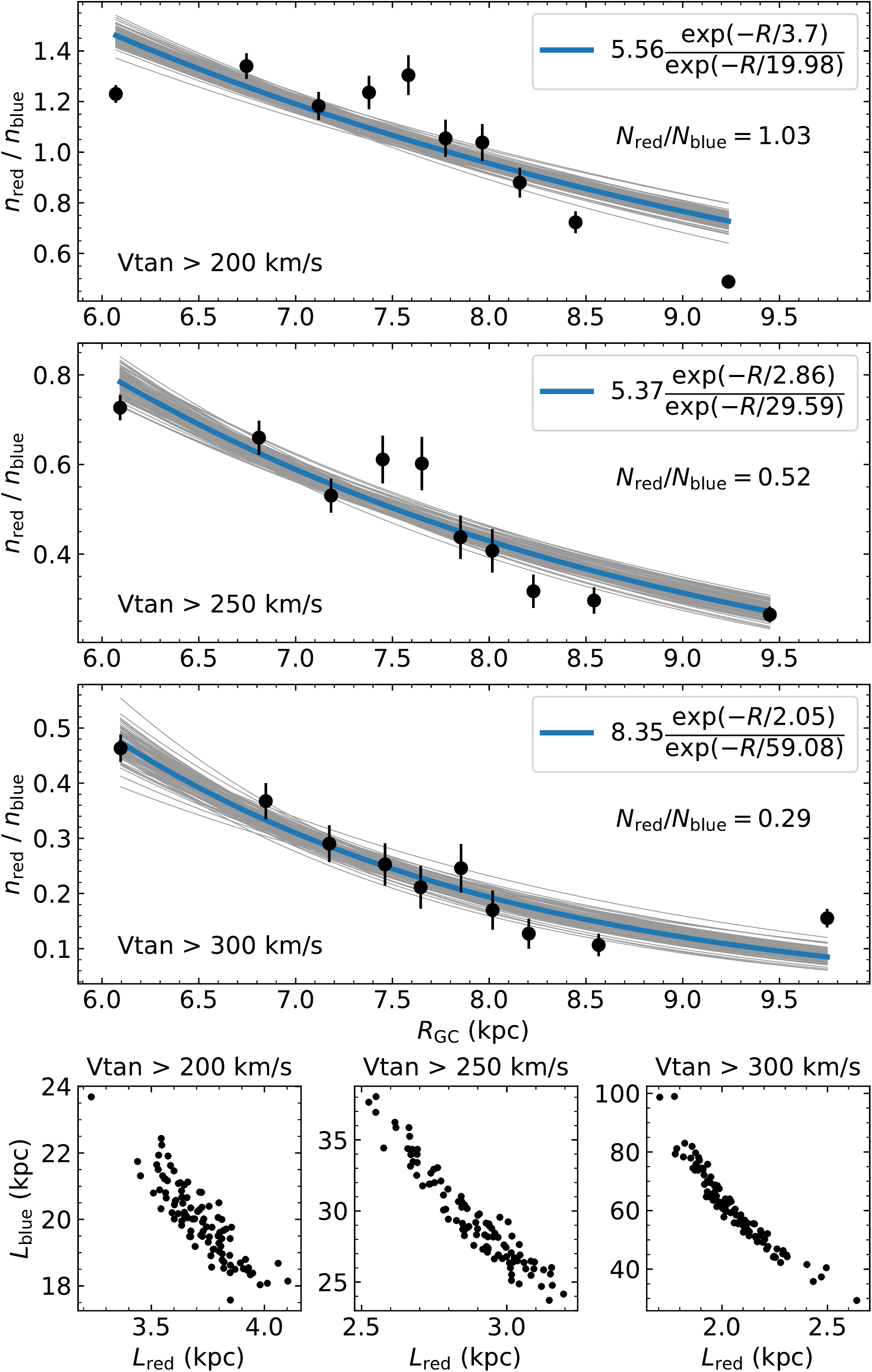}
\caption{Relative number density of red-to-blue sequence {\it Gaia} stars as a function  of galactocentric radius for three different cuts in tangential velocity (upper three panels).
The data have been divided into 10 bins of equal numbers of stars, and the uncertainties are based on 1000 bootstrap samples.
The blue curve is a fit of the data to Eq.~\eqref{eq:density_frac}, and the gray curves are a subset of the fits to individual bootstrap samples.
The lower three panels show the scale lengths from the fits to each of the bootstrap samples.}
\label{fig:red_blue_number_density}
\end{figure}

The position of the red-sequence stars in Fig.~\ref{fig:vtan_feh}, as the tip of a population extending to high metallicities, indicates that they are the high-velocity tail of the underlying disk population.
The somewhat arbitrary definition of high-velocity stars as those with $V_{\rm T} > 200$ km\,s$^{-1}$ makes the blue and red sequences equally populated in the H-R~diagram.
If we make the cut at $300$ km\,s$^{-1}$ instead, the red sequence almost completely disappears from the H-R~diagram, emphasizing the distinct natures of the two populations.

To further test the link between the red sequence and the thick disk, we take all of the {\it Gaia} DR2 high-velocity stars (applying the same quality cuts as previously) and investigate the distribution of stars between the two sequences as a function of distance, $R_{\mathrm{GC}}$, from the galactic center.
$R_{\mathrm{GC}}$ has been calculated based on {\it Gaia} astrometry, assuming the Sun is $8$~kpc from the Galactic center \citep{2019A&A...625L..10G}.
In Fig.~\ref{fig:hrd3} the high-velocity stars have been divided into three bins in galactocentric radius.
The stars in the solar annulus are split approximately equally between the blue and red sequences when counting stars on the RGB (within $\pm 0.05$~mag in color from the two isochrones in Fig.~\ref{fig:hrd3}).
In the inner Galaxy the red sequence has more stars with the fraction of red-to-blue sequence stars being 1.25.
Moving to the outer Galaxy, the blue sequence dominates with about two times the number of red-sequence stars.
Thus, assuming that the selection functions are the same for stars in each sequence (they have been selected in the same absolute magnitude range) the red-sequence population is more centrally concentrated than the blue sequence.
This is consistent with our expectation if the red sequence is a part of the chemically defined thick disk, which is known to be absent in the outer Galaxy due to its relatively short scale-length \citep[e.g.,][]{2011ApJ...735L..46B}.

To further quantify the radial distributions of stars in the two sequences, the data have been divided into 10 bins in $R_{\mathrm{GC}}$ with equal numbers of stars, and in each bin the numbers of stars belonging to the red and blue sequences have been counted on the RGB.
In Fig.~\ref{fig:red_blue_number_density}, the relative number of red-to-blue sequence stars is shown for each of the 10 bins in $R_{\mathrm{GC}}$ for all high-velocity stars and for two stricter cuts on the tangential velocity.
As expected based on the data in Fig.~\ref{fig:hrd3}, this fraction decreases with increasing $R_{\mathrm{GC}}$.

We can attempt to use these relative number densities to estimate the scale lengths of the two populations, $L_{\mathrm{red}}$ and $L_{\mathrm{blue}}$, starting from the assumption that both populations follow an exponential number density distribution given by
\begin{align}
    n(R) = k\exp\left(-\frac{R}{L}\right) \, .
\end{align}
The constant $k$ is determined by the total number $N$ of stars in the population
\begin{align}
    N &= \int_0^{\infty}k\exp\left(-\frac{R}{L}\right) \mathrm{d}R \nonumber \\
    &= k\left[-L\exp\left(-\frac{R}{L}\right)\right]_0^{\infty} \\
    &= kL \, , \nonumber
\end{align}
giving $k=N/L$.
The ratio of two such distributions is given by
\begin{align} \label{eq:density_frac}
    \frac{n_{1}}{n_{2}} = \frac{N_{1}}{N_{2}}\frac{L_{2}}{L_{1}}\frac{\exp(-R/L_{1})}{\exp(-R/L_{2})} \, .
\end{align}
This expression has been fitted to the data in Fig.~\ref{fig:red_blue_number_density} with $L_{1}$ and $L_{2}$ allowed to vary freely and $N_{1}/N_{2}$ fixed at the observed ratio.

The exponential fit does not describe the data very well when we include all high-velocity stars ($V_{\rm T} > 200$ km\,s$^{-1}$; Fig.~\ref{fig:red_blue_number_density} upper panel).
Instead, the number densities are roughly equal until about 7.5~kpc after which the ratio drops off.
When making the velocity cut stricter, the exponential profile fits better.
The best-fitting values for $L_{\mathrm{red}}$ are $2.86$~kpc ($V_{\rm T} > 250$ km\,s$^{-1}$) and $2.05$~kpc ($V_{\rm T} > 300$ km\,s$^{-1}$), and for $L_{\mathrm{blue}}$ they are about $30$ and $60$~kpc.
As shown in the lower panels of Fig.~\ref{fig:red_blue_number_density}, the two scale lengths are strongly anticorrelated in the fit, and $L_{\mathrm{red}}$ varies within a range of about $0.5$~kpc.

\subsection{Comparison with Nissen \& Schuster halo stars}
\label{sect:NS}

\begin{figure}
\centering
\includegraphics[width=\columnwidth]{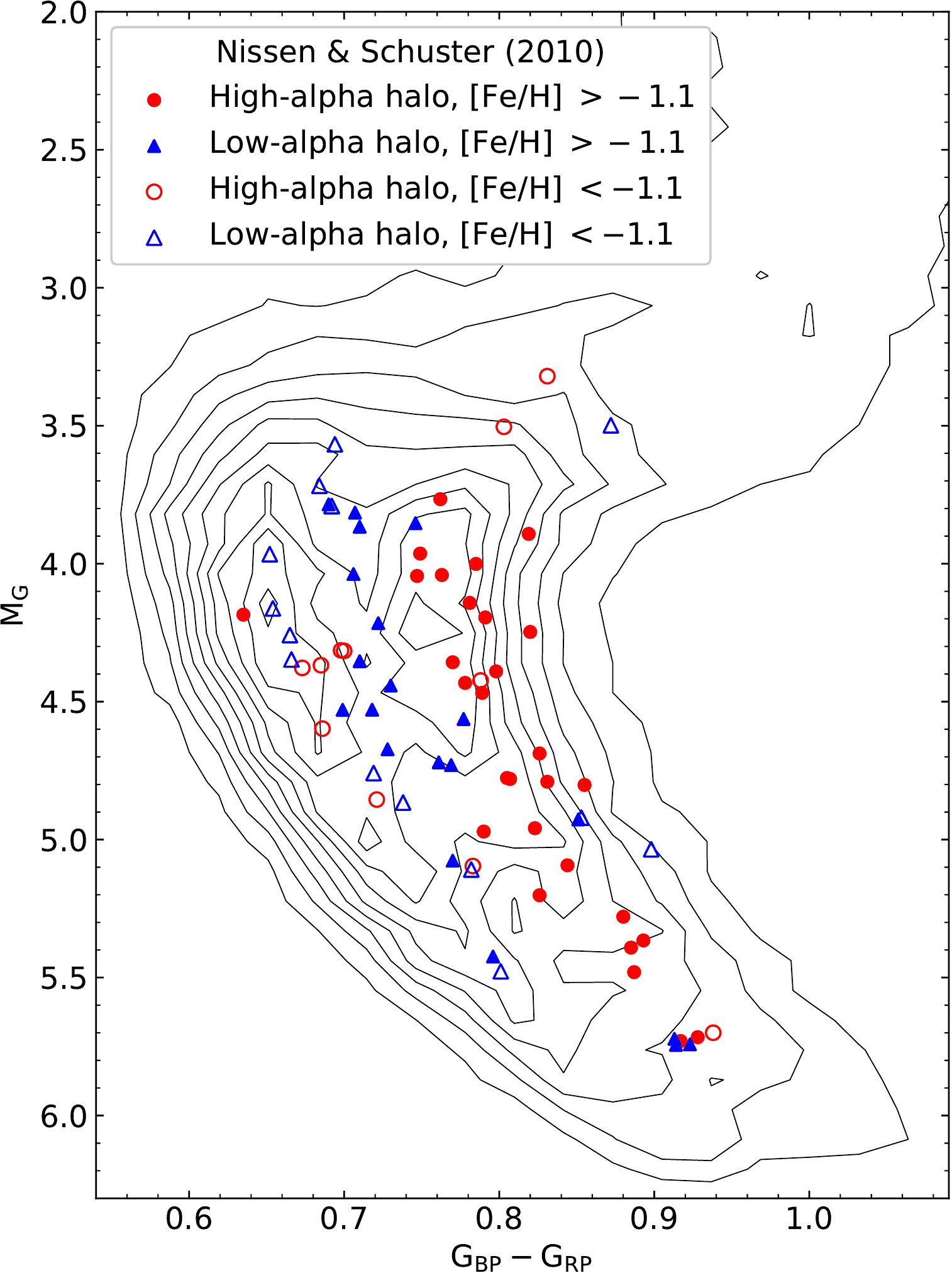}
\caption{\textit{Gaia} colors and magnitudes for the \citet{2010A&A...511L..10N} high- and low-$\alpha$ stars on top of density contours of the high-velocity SkyMapper stars.}
\label{fig:ns}
\end{figure}

\citet{2010A&A...511L..10N} studied 94 dwarf stars with halo and thick-disk kinematics (of the 94 stars, 16 have thick-disk kinematics).
The stars were selected based on their metallicities (derived from Str\"omgren photometry) and their total space velocities (for the halo this meant $V_{\rm Tot, LSR}> 180$\,km\,s$^{-1}$).
When the high-resolution, high-signal-to-noise spectra for the selected stars were analyzed, the stars showed an unexpected pattern in elemental abundances: they split into two sequences.
One with high-$\alpha$ and one with low-$\alpha$. In addition, the low-$\alpha$ sequence also has low [Ni/Fe], which is unusual. 
It has been suggested that these two halo populations are related to the two sequences we see in the high-velocity sample studied here \citep{2018A&A...616A..10G, 2018ApJ...863..113H}. 

Fig.~\ref{fig:ns} shows a zoom in around the turnoff  of the \text{Gaia} H-R diagram.
In this H-R diagram we have also plotted the high-$\alpha$ (red circles) and the low-$\alpha$ (blue triangles) halo stars from \citet{2010A&A...511L..10N}.
These stars show two clear trends in elemental abundance space.
Here the stars with [Fe/H]~$<-1.1$ (open symbols) mainly lie along the blue sequence regardless of their $\alpha$-enhancement.
The stars with [Fe/H]~$>-1.1$ (filled symbols), however, split up with the high-$\alpha$ stars along the red sequence and the low-$\alpha$ stars in between the density peaks of the two sequences.
Hence, if we use the original classifications of \citet{2010A&A...511L..10N} based solely on the elemental abundance ratios then it is clear that the two chemically defined trends do not align with the two sequences seen in the high-velocity \textit{Gaia} H-R diagram.
If, instead, the stars are divided according to metallicity, as also done by \citet{2018ApJ...863..113H}, then stars of progressively lower and lower metallicities appear more and more to the left in the H-R diagram.
This further testifies to the fact that the two sequences found in the high-velocity sample are mainly driven by metallicity.

\subsection{Comparison with MDFs from Gallart et al.}
\label{sect:gallart}

\begin{figure}
\centering
\includegraphics[width=\columnwidth]{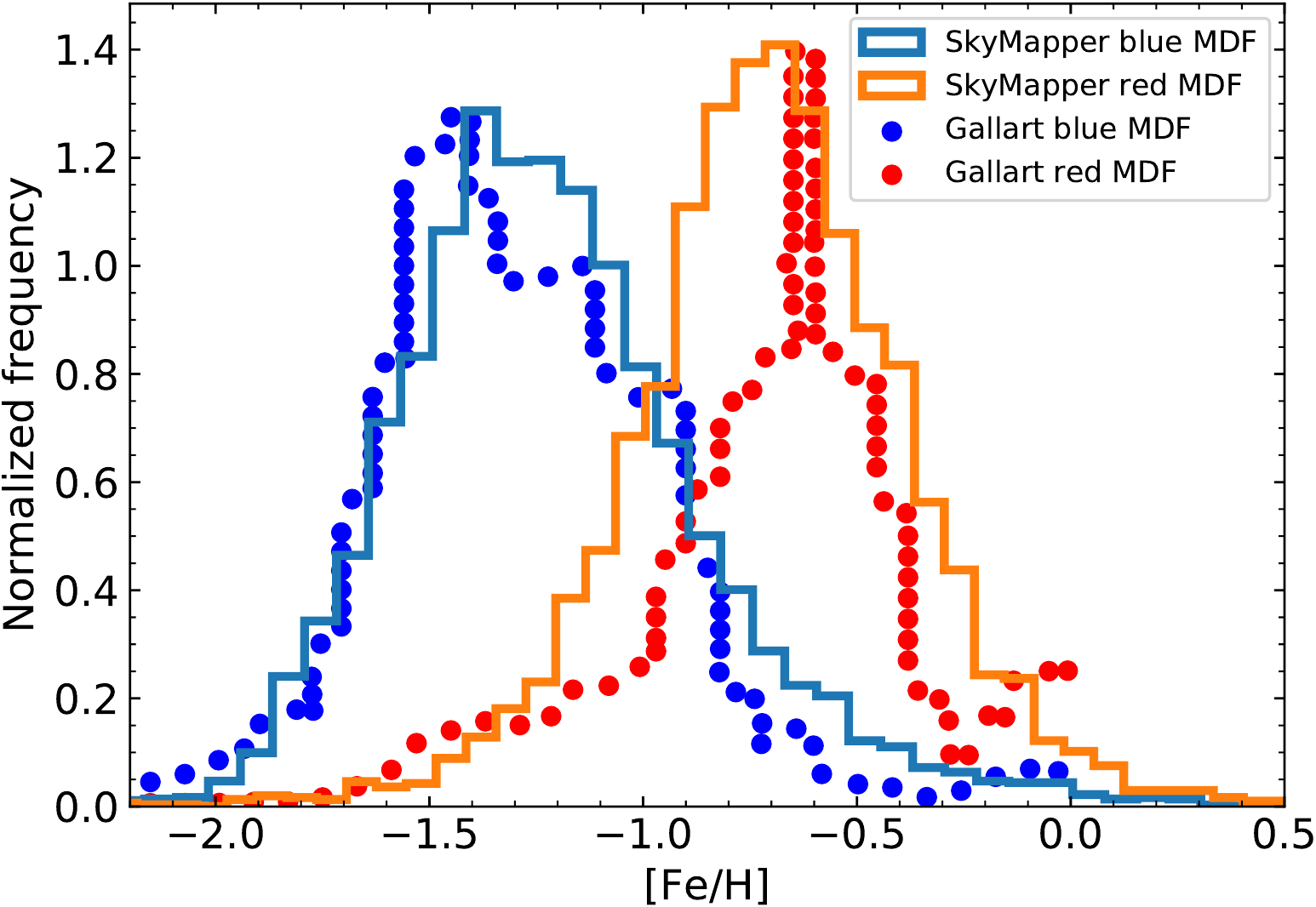}
\caption{Comparison of the MDFs derived for the blue and red sequences in this work and those by \citet{2019arXiv190102900G} reproduced from their Figure~2 (see the legend).
The \citet{2019arXiv190102900G} MDFs have been rescaled to match the maxima of the SkyMapper MDFs.}
\label{fig:compg}
\end{figure}

\citet{2019arXiv190102900G} use the \textit{Gaia} photometry to model the full H-R diagram for the high-velocity stars. They find that they can faithfully reproduce the H-R diagram using a range of (old) ages and metallicities.
We are now in a position to check if the MDFs they derive based on their H-R diagram fitting are consistent with a much more comprehensive data set of [Fe/H].

In Fig.~\ref{fig:compg} we compare our resulting MDFs for the red and the blue sequences with those found by \citet{2019arXiv190102900G} (here we concentrate on the RGB that has the best metallicity estimates; see Sect.~\ref{sect:params}).
We find that \citet{2019arXiv190102900G} and the SkyMapper-based MDF for the blue sequence agree well.
However, the comparison of the red sequence is less favorable, while the SkyMapper MDF is broad the MDF resulting from the H-R diagram fitting is very narrow, centered at $-0.6$~dex.
We note that with our simple cut in the RGB (see Fig.~\ref{fig:hrd1}) it is entirely possible that we misclassify some stars as being on the red sequence whilst really being on the blue and vice versa.
The problem is that the contamination works both ways; in our case if we have classified blue-sequence stars as red, then we should see an even broader blue MDF.
This would not agree well with those from \citet{2019arXiv190102900G}.
Hence, it does not seem easy to reconcile the SkyMapper red-sequence MDF with that by \citet{2019arXiv190102900G}.

\section{Discussion and conclusions}
\label{sect:disc}

Using the so far largest data set with well-determined metallicities we have confirmed and extended the findings that the high-velocity stars in \textit{Gaia} DR2 show two distinct stellar populations.
In particular, we show using some 10,000 high-velocity RGB stars that the two stellar sequences are driven by a difference in metallicity with the blue and red sequences peaking at $[\mathrm{Fe}/\mathrm{H}]=-1.4$ and $-0.7$~dex, respectively.
Matching isochrones with metallicities around these peaks to the turnoffs in the H-R diagram, we find that the sequences are made up of stars with the same range of ages ($\gtrsim 10$~Gyr).
Although the peaks of the two sequences are rather well defined, there is a metallicity spread within each sequence.
Hence, slightly different combinations of lower (higher) ages with higher (lower) metallicities equally fit the data along each sequence, and this naturally explains the sharp turnoffs and broader RGBs in the H-R diagram (Fig.~\ref{fig:hrd2}).

Our results thus show that the two populations formed stars at a similar time, which, considering the rather different metallicities, further strengthens the claim that at least one of them is the result of a merger event.
This is in line with the conclusion reached in previous studies of the stars making up the blue sequence \citep{2018arXiv181208232D, 2018ApJ...863..113H, 2018Natur.563...85H}.
It would be natural to conclude that the blue sequence is the merging population as it has the lower metallicity and hence could be the result of star formation in a less massive body whilst the red sequence is more metal-rich, which requires a heavier body to reach the higher metallicities \citep[compare, e.g., ][]{2004MNRAS.351.1338L}.
The merger likely took place after all the stars formed -- which leaves a lower limit of about 10\,Gyr ago for the merger to have happened \citep[similar to what is concluded in][]{2018arXiv181208232D, 2018Natur.563...85H}.
This is consistent with observational evidence of the quiescent evolution (i.e. no major disturbances) of the Milky Way disk over the past 8--10~Gyr \citep{2015MNRAS.450.2874R, 2016MNRAS.455..987C}.

There is some overlap between the stars of the two sequences in the $V_{\rm T} - [\mathrm{Fe}/\mathrm{H}]$ plane, but they clearly display different distributions.
The red sequence appears as a continuation of the underlying disk population in which the mean $V_{\rm T}$ increases steadily with decreasing metallicity.
At the lowest metallicities the number of stars above the somewhat arbitrary cut of $V_{\rm T} = 200$~km\,s$^{-1}$ is big enough for the population to stand out in the H-R~diagram.
The blue sequence, on the other hand, shows little to no variation of the mean $V_{\rm T}$ across the metallicity range where it dominates and is clearly separate from the disk population.

We find that the red-sequence population is more centrally concentrated in the Galaxy than the blue sequence.
This result is meaningful only if the same selection function applies to stars of the two sequences, which seems to be a reasonable assumption given that the stars are selected within the same range of absolute magnitudes.
With the additional assumption that both populations can be described by an exponential radial density distribution, we find a scale length of the red-sequence stars of about 2--3~kpc.
This range is consistent with the estimated scale length of about $2\pm0.2$~kpc for the chemically defined thick disk \citep{2016ARA&A..54..529B}.
Again, these conclusions are in line with previous studies in which the red sequence has been interpreted as the high-velocity tail of the thick disk \citep{2018arXiv181208232D, 2018ApJ...863..113H, 2018Natur.563...85H}.
The estimated blue-sequence scale length is greater than $20$~kpc which implies that the stellar density varies very little with radius in the observed region.

The apparent discrepancy between the interpretation laid out here and the one presented by \citet{2019arXiv190102900G}, in which the red sequence is identified as the in-situ halo, may simply come down to different definitions of the populations.
\citet{2019arXiv190102900G} define halo stars as those with $V_{\rm T} > 200$~km\,s$^{-1}$; however, as we argue in Sect.~\ref{sect:thick_disk}, this definition is arbitrary and we see no clear separation between the red-sequence stars and the underlying disk population.
This difference in definitions cannot, however, explain the difference we find between the SkyMapper and their red-sequence MDF since we use the same cut of $V_{\rm T} > 200$~km\,s$^{-1}$.
Instead, this difference may be explained by the details of their H-R diagram modeling, e.g., inaccuracies in the stellar models.

All things considered, our results corroborate the picture in which the two sequences are made up of metal-poor (likely accreted) halo stars and (relatively) more metal-rich stars from the tail of the thick disk.
Notably, the position of the red-sequence stars in the $V_{\rm T} - [\mathrm{Fe}/\mathrm{H}]$ plane, and their spatial distribution, has strengthened the link between the red sequence and the Galactic disk.

\acknowledgments
C.L.S. and S.F. were supported by the project grant The New Milky Way from the Knut and Alice Wallenberg foundation and by the grant 2016-03412 from the Swedish Research Council.
L.C. is the recipient of the ARC Future Fellowship FT160100402.
This work has made use of data from the European Space Agency (ESA) mission {\it Gaia} (\url{https://www.cosmos.esa.int/gaia}), processed by the {\it Gaia} Data Processing and Analysis Consortium (DPAC, \url{https://www.cosmos.esa.int/web/gaia/dpac/consortium}).
Funding for the DPAC has been provided by national institutions, in particular the institutions participating in the {\it Gaia} Multilateral Agreement.

%

\vspace{5mm}
\facilities{SkyMapper, {\it Gaia}}

\vspace{-3mm}


\end{document}